\documentclass[11pt]{article} \usepackage{amssymb}
\usepackage{amsfonts} \usepackage{amsmath} \usepackage{amsthm} \usepackage{bm}
\usepackage{latexsym} \usepackage{epsfig}

\newtheorem*{theorem*}{Theorem}
\newtheorem{theorem}{Theorem}[section]
\newtheorem{lemma}[theorem]{Lemma}
\newtheorem*{proposition*}{Proposition}

\newtheorem{proposition}[theorem]{Proposition}
\newtheorem{mechanism}[theorem]{Mechanism}

\newtheorem{definition}[theorem]{Definition}

\newcommand{\ignore}[1]{}

\newcommand{\enote}[1]{} \newcommand{\knote}[1]{}
\newcommand{\rnote}[1]{}

  \renewcommand{\P}{{\bf P}} 
\newcommand{\E}{{\bf E}}

\newcommand{\eps}{\epsilon}

 \newcommand{\R}{\mathbb R}

\newcommand{\half}{{\textstyle \frac12}}

\renewcommand{\phi}{\varphi}

\begin{document}
\title{Truthful Fair Division}
\author{ Elchanan Mossel\footnote{Weizmann Institute and
    U.C. Berkeley. E-mail: mossel@stat.berkeley.edu. Supported by a
    Sloan fellowship in Mathematics, by BSF grant 2004105, by NSF
    Career Award (DMS 054829) by ONR award N00014-07-1-0506 and by ISF
    grant 1300/08}~ and Omer Tamuz\footnote{Weizmann
    Institute. E-mail: omer.tamuz@weizmann.ac.il. Supported by ISF
    grant 1300/08} }

\maketitle
\begin{abstract}
  We address the problem of fair division, or cake cutting, with the
  goal of finding truthful mechanisms. In the case of a general
  measure space (``cake'') and non-atomic, additive individual
  preference measures - or utilities - we show that there exists a
  truthful ``mechanism'' which ensures that each of the $k$ players
  gets at least $1/k$ of the cake. This mechanism also minimizes risk
  for truthful players.  Furthermore, in the case where there exist at
  least two different measures we present a different truthful
  mechanism which ensures that each of the players gets more than
  $1/k$ of the cake.

  We then turn our attention to partitions of indivisible goods with
  bounded utilities and a large number of goods. Here we provide
  similar mechanisms, but with slightly weaker guarantees.  These
  guarantees converge to those obtained in the non-atomic case as the
  number of goods goes to infinity.

\end{abstract}

\section{Introduction}
The basic setting of fair partition problems includes a ``cake'' -
some divisible resource - and a number of players, each with different
preferences with regards to the different ``pieces'' of the cake. The
task is to divide the cake among the players in a way that would be
``fair'' according to some fairness notion. This applies to numerous
situations, from divorce trials to nuclear arms
reduction, see e.g.~\cite{BramsTaylor:96}.

This problem has been studied extensively both in the non-atomic case
and in the case of indivisible goods. In this paper we take a fresh
look at fair division, by approaching it from the mechanism design
point of view, and in particular in search of truthful partition
mechanisms.

\subsection{Background}
The problem of fair division, or cake cutting, is a central and
classical problem in economics.  A {\em fair} partition is one where
each of $k$ players receives at least $1/k$ of the cake, each
according to her own measure (this is also known as a {\em
  proportional} partition).  Fair partition into two parts has long
been known to be possible using a ``cut-and-choose'' procedure: a cake
is guaranteed to be fairly divided in two if one player cuts it and
the other picks a piece.

Fair partition to more than two players requires non-trivial
mathematics.  The founding work in this field was done by Steinhaus,
Knaster and Banach in the forties (e.g. \cite{Knaster:44},
\cite{Steinhaus:48}, \cite{Steinhaus:49}). They and others (e.g.
\cite{Neyman:46}) proved existence theorems for fair division on
general measure spaces and non-atomic measures, where ``fairness'' was
again, in general, taken to mean that each of $k$ players receives at
least $1/k$ of the cake, each according to her own measure. These
proofs are not constructive, but nevertheless were useful in
generalizing ``cut-and-choose'' to more than two players.

Another natural approach is ``moving knife algorithms'', first
described by Dubins and Spanier \cite{DubinsSpanier:61}. These are not
algorithms in the modern Turing Machine sense of the word (as
``cut-and-choose'' and its generalizations aren't), but still provide
a ``practical'' way to cut actual cakes, if nothing else, by the same
fairness criterion mentioned above.

A stronger concept of fairness is that of ``envy-free'' partitions
(\cite{Foley:67},\cite{Varian:74}).  In an envy free partition there
is no player who would trade her allocation with one given to another
player. Such partitions have been studied extensively (e.g., see
\cite{BramsTaylor:96}).

Partitioning indivisible goods is a more recently studied variant,
which places the problem in a standard algorithmic setup. However, it
is easy to see that fairness or envy-freeness could not be achieved in
every setup as the example of one good demonstrates. Still, the
results of Lipton et al.~\cite{Lipton:04} show that almost envy free
partitions exist. These are partitions where no player envies another
player by more than the value of a single good.

Classical cake cutting mechanisms such as ``cut-and-choose'' usually
require the players to choose a piece of a subset of the cake or make
a cut in it, according to their preferences. Alternatively, the
mechanism queries the player about her valuation of a particular piece
of the cake. We take an approach which is more prevalent in the modern
mechanism design world: each player declares their preference (i.e.,
their entire measure on the cake) to some ``third party'', which then
proceeds to divide the cake according to a predetermined algorithm.

While circulating a draft of this paper, it was brought to our
attention that similar questions are discusses in a recent working
paper by Chen et al.\ \cite{Chen:10}.  They restrict themselves to a
particular class of measures: ``the case where the agents hold
piecewise uniform valuation functions, that is, each agent is
interested in a collection of subintervals of $[0, 1]$ with the same
marginal value for each fractional piece in each subinterval.'' In
this setting, they show a truthful, deterministic, polynomial-time,
fair and envy-free mechanism. For the more general setting which we
study they independently derive our mechanism~\ref{mech:k}.

\subsubsection{Truthfulness}
The notion of truthfulness is very natural in partitioning
problems. Why would a player declare her true measure if declaring a
different measure would result in a better partition for her?

The ``cut-and-choose'' method is in some sense truthful: the players
don't have to trust each other to be guaranteed a half of the
cake. However, it is {\em not} truthful in the sense that players have
incentive to ``strategize'' in order to increase the value of the
piece they receive. For example, assume that the players' preferences
differ, and that the first player (the ``cutter'') knows, or guesses,
the second player's (the ``chooser'') preference. She may then cut the
cake into two pieces such that the first piece is worth much more than
one half to herself, and $\half-\epsilon$ to the chooser. This would,
perhaps, secure her a large piece of the cake, and leave the chooser
with the feeling that dealings were not completely {\em fair}. To
further confound matters we note that the chooser may realize, in this
setting, that she could gain a larger piece by manipulating the
cutter's perception of her preferences.

As noted in~\cite{BramsTaylor:96}, many of the partition mechanisms
which were discovered in the non-atomic case have this same weaker
truthfulness property; for fair partitions - they have the property
that truthful players are guaranteed to receive at least $1/k$ of their
total value of the cake. Similarly for envy free partitions every
truthful player is guaranteed not to envy any of the other players.

Still the question remains as to why is it beneficial for a player to
declare her true value?

A work which addresses this question is a paper by Lipton et
al.\ \cite{Lipton:04} who analyze a truthful mechanism for allocating a
set of indivisible goods: simply give each good to each of $k$ players
with probability $1/k$. This mechanism is further analyzed in
\cite{Caragiannis:09}, who showed that this partition results, with
high probability, in $O(\alpha\sqrt{n\ln k})$ envy, where $n$ is the
number of goods, $k$ is the number of players, and $\alpha$ is the
maximum utility over all goods and players.

The strongest possible notion of truthfulness is the following: a
mechanism is {\em truthful} if it is always the case that a player's
utility when declaring the truth is as high as her utility when lying.
The notion of~\cite{Lipton:04,Caragiannis:09} is weaker. They focus,
as we do, on {\em truthfulness in expectation}: the expected value of
a player's utility is maximal when telling the truth.


\subsection{Our Results}

\subsubsection{Non Atomic Measures}
Although our ``third party'' approach is one of modern mechanism
design, our work in the non-atomic setup lies in the realm of
classical fair division, in that it does not consider the
computational aspects of the mechanisms: in some cases, an infinite
amount of information would have to be conveyed to the third
party. Even when not, the calculations required may be intractable or
even not recursive.

We first consider the problem of a truthful fair partition with
non-atomic measures.  For this problem, using a result of Steinhaus
\cite{Steinhaus:49}, we show that such a partition always exists.
Furthermore, our mechanism particularly incentivizes risk averse
players to play truthfully.

We next ask if it is possible to devise a truthful mechanism that
guarantees that each player gets more than $1/k$ of the
cake. Obviously the answer is negative if all the measures are the
same. However, as mentioned by Steinhaus \cite{Steinhaus:48} and
proved by Dubins and Spanier \cite{DubinsSpanier:61}, if the measures
are not all identical, then there exists a partition where each player
gets more than $1/k$ of the entire cake.

Our main result is a {\em truthful in expectation} mechanism which
guarantees that each player gets at least $1/k$ of the cake, and
furthermore the expected size of the piece each player gets is
strictly larger than $1/k$.

We further show that there exists no deterministic and truthful
mechanism that gives these guarantees and so only in the weaker notion
of truthfulness it is possible to obtain such results.

An additional argument in favor of randomized mechanisms is that a
deterministic mechanism cannot be symmetric: when all the players
declare the same preferences, it must arbitrarily break the symmetry
and assign different players different slices of the cake. A
randomized algorithm can avoid this.

\subsubsection{Indivisible Goods}
Our results presented in the case of non-atomic measures are
existential; we do not provide protocols for implementing them. To
address the computational aspect of the problem we consider partitions
of a large number of indivisible goods where the number of goods $n$
is exponential in the number of players, and the utility of a single
good is bounded. We give efficient versions of all of our mechanisms
in this setup, with guarantees that are slightly weaker than those
provided in the continuous case. The guarantees hold in expectation,
and moreover there is a deterministic guarantee for each player to
receive a share of the goods which is at least $(1-\eps)$ of the
expected value, where $\eps = O(M^k/n)$ where $k$ is the number of
players, $n$ is the number of goods and $M$ is the maximum utility of
a single good.

More generally, we prove a discrete analogue to a theorem of Dubins
and Spanier \cite{DubinsSpanier:61}. They show that the space of
partition utilities is convex. We show that the same is true in the
discrete case, again up to a factor of $(1-\epsilon)$, as defined
above.

\section{Continuous Truthful Mechanisms}
\subsection{Existence Theorems for Fair Divisions}
Dubins and Spanier \cite{DubinsSpanier:61}, rephrasing Fisher
\cite{Fisher:38}, provide the following description of what they call
``The problem of the Nile'':

``Each year the Nile would flood, thereby irrigating or perhaps
devastating parts of the agricultural land of a predynastic Egyptian
village. The value of different portions of the land would depend upon
the height of the flood. In question was the possibility of giving to
each of the $k$ residents a piece of land whose value would be $1/k$
of the total land value no matter what the height of the flood.''

Neyman \cite{Neyman:46} showed that this is possible, given that there
are a finite number of levels that the Nile can rise to.

Let $\mathcal{C}$ be a ``cake'' (a set), and $\mathfrak{C}$ a set of
``slices'' (a $\sigma$-algebra of subsets of $\mathcal{C}$). Let there
be $k$ players, and let $\mu_1,\ldots,\mu_k$ be non-atomic probability
(additive) measures on $(\mathcal{C},\mathfrak{C})$, so that the value
of a slice $C \in \mathfrak{C}$ to player $i$ is $\mu_i(C)$. Then
Neyman's theorem establishes that there exists a partition of the cake
$C_1,\ldots,C_k$ such that for all players $i$ and slices $j$ it holds
that $\mu_i(C_j)=1/k$. Hence all the slices are equal, by all the
player's measures.

Dubins and Spanier \cite{DubinsSpanier:61} show that a better
partition is always possible when at least two of the players have
different measures. Their theorem implies that in this case a
partition is possible for which, for all players and slices $i$, it
holds that $\mu_i(C_i)>1/k$, and so each player gets strictly more
than $1/k$ of the cake, by his or her own measure.

\subsection{Truthful Mechanisms}
\subsubsection{Fair division}
We present a simple truthful ``mechanism'' for distributing the cake
among $k$ players, which assures that each player gets precisely $1/k$
of the cake, by all the players' measures. It is a ``mechanism'' in
quotes since it is as constructive as Neyman's theorem, which is not
constructive. Note that this mechanism also appears in~\cite{Chen:10}.

\begin{mechanism}
  \label{mech:k}
  Assume the players' true measures are $\mu_1,\ldots,\mu_k$, and that
  they each declare some measure $\nu_i$.  Find a partition
  $C_1,\ldots,C_k$ such that $\forall i,j:\nu_i(C_j)=1/k$. Then choose
  a random permutation $\tau$ of size $k$, from the uniform
  distribution, and give $C_{\tau(i)}$ to player $i$.
\end{mechanism}

\begin{proposition}
  \label{prop:k}
  Mechanism~\ref{mech:k} is {\em truthful} in the following sense: No
  player can increase her expected utility by playing non-truthfully.
  Further, a player who plays truthfully minimizes the risk/variance
  of the measure of the piece she gets.
\end{proposition}

\begin{proof}
  The expected size of the slice for player $i$ is
  $\sum_j\mu_i(C_j)\P[\tau(i)=j]=\sum_j\mu_i(C_j)/k=\mu_i(\cup
  C_j)/k=1/k$. Since it is independent of $\nu_i$ then player $i$ has
  no incentive to declare untruthfully. Furthermore, a player that
  declares $\nu_i=\mu_i$ is guaranteed a slice of size $1/k$, and so
  the truth minimizes the variance (or risk), to zero.
\end{proof}

\subsubsection{Super-fair division}
For this result we set $\mathcal{C}=[0,1) \in \R$ and let
$\mathfrak{C}$ be the Borel $\sigma$-algebra. While this result can be
extended to more general classes of spaces and algebras, we present it
in this restricted form for clarity. We consider the case where at
least one pair of measures are not equal, i.e. the case in which
``super-fair'' partitions exist --- partitions in which
$\mu_i(C_i)>1/k$.

We first provide motivation for our usage of ``truthfulness in
expectation'', by showing that deterministic ``super-fair'' mechanisms
cannot be truthful:
\begin{theorem}
  \label{thm:no_determinism}
  Any deterministic mechanism that gives each player $1/k$ of the cake
  when all declared measures are equal and more than $1/k$ of the
  cake otherwise is not truthful.
\end{theorem}
\begin{proof}
  Consider the case where all players declare the same arbitrary
  measure $\mu$. Then they receive slices $C_i$ such that $\forall
  i:\mu(C_i)=1/k$. Now, consider the case where player 1's true
  measure $\nu$ is such that $\nu(C_1)=1$. Then player 1's utility for
  declaring $\mu$ is 1. We propose that her utility for declaring $\nu$
  (i.e., being truthful) is less than one, and therefore the mechanism
  is not truthful.

  Assume by way of contradiction that her utility for declaring $\nu$
  is one. Then in this case she must also receive slice $C_1$, since
  that is the only slice worth one to her. But if player 1 receives
  $C_1$ then the rest of the players have, by their measure $\mu$,
  exactly $(k-1)/k$ of the cake left to share, and so it is impossible
  that they all receive more than $1/k$ of it. This contradicts the
  hypothesis, since $\mu(C_1)=1/k$ and $\nu(C_1)=1$, and therefore
  $\mu \neq \nu$ and all players must receive more than $1/k$.
\end{proof}

We now describe a randomized mechanism that is ``super-fair'' and
truthful in expectation.

\begin{mechanism}
  \label{mech:kepsilon}
  Assume again that $\mu_1,\ldots,\mu_k$ are the players' true
  measures, and that they each declare some measure $\nu_i$. To
  distribute the cake, pick a partition $C_1,\ldots,C_k$ from a
  distribution $D$ over partitions, which we describe below. If it so
  happens that $\nu_i(C_i)>1/k$ for all $i$, then distribute the
  slices accordingly. Otherwise distribute by mechanism~\ref{mech:k},
  that is, give a slice of value $1/k$ to all players.
\end{mechanism}
\begin{proposition}
  \label{prop:k_eps_works}
  In mechanism~\ref{mech:kepsilon} the expected utility of a truthful
  player is larger than $1/k$ if super-fair partitions are picked with
  positive probability.
\end{proposition}
\begin{proof}
  If  $C_1,\ldots,C_k$ is super-fair and the players are truthful, then this
  partition is recognized as super-fair, and the players each get strictly more
  than $1/k$.
  In the event that the picked partition is not super-fair, and the players
  are truthful, then the mechanism reverts to giving the players precisely
  $1/k$ of the cake. Thus the expectation for truthful players is more
  than $1/k$.
\end{proof}
\begin{proposition}
  \label{prop:k_eps_truth}
  In mechanism~\ref{mech:kepsilon} playing truthfully maximizes a
  player's expected utility.
\end{proposition}
\begin{proof}
  Consider again two cases: the first, in which  $C_1,\ldots,C_k$ is
  super-fair, and the second, in which it isn't.

  In the first event, a truthful player's expected share is
  more than $1/k$. Playing untruthfully could either have no effect, leaving
  the utility as is, or else the only other possibility is that the partition
  is misconstrued not to be super-fair, in which case the player's
  utility is reduced to $1/k$.

  In the second event, in which the partition picked is not super-fair,
  playing untruthfully may again have no effect, leaving the utility at
  $1/k$. However, if, to some player, the share allocated by this partition
  was worth less than $1/k$, playing untruthfully may make it seem to be
  valued at more then $1/k$, turning the partition into super-fair by the
  declared preferences, and resulting in a utility less than $1/k$ for that
  player.

  Thus, for any random choice of $C_1,\ldots,C_k$ the truthful
  player's expected utility is maximal, and the proposition follows.
\end{proof}

To assure that this mechanism results in a slice of expected size strictly
greater than $1/k$, we must find a distribution $D$ (from which we draw
the partition) such that for any set of measures, where at least one pair
is not equal, with positive probability
$\mu_i(C_i)>1/k$ for all $i$. To this end we make the following definition:

\begin{definition}
  Denote by $\mathcal{Q}$ the set
  of partitions $C_1,\ldots,C_k$ of $[0,1) \in \R$ for which each $C_i$ is a
  finite union of half-open intervals with rational endpoints.
\end{definition}
We note that  $\mathcal{Q}$ is countable. $D$ now need only be some distribution with
support $\mathcal{Q}$:
\begin{theorem}
  \label{thm:super_fair}
  Let $\mu_1,\ldots,\mu_k$ be non-atomic probability measures on
  $[0,1) \in \R$ and the Borel $\sigma$-algebra,
  such that there exist $i,j$ for which $\mu_i \neq \mu_j$. Let $D$
  be a distribution over the partitions $C_1,\ldots,C_k$ of $[0,1)$ into $k$
  sets, such that the support of $D$ is $\mathcal{Q}$. Then
  \begin{equation*}
    \P_D[\forall i:\:\mu_i(C_i)>1/k]>0.
  \end{equation*}
\end{theorem}

The proof appears in Appendix~\ref{app:kepsilon_proof}.

\section{Indivisible Goods}

\label{section:model}
Let  $\mathcal{C}=\{a_1,\ldots,a_n\}$ be a finite set of indivisible goods (``discrete cake'').
Let there be $k$ players, where each
has an additive bounded measure (utility) on the algebra of subsets of
$\mathcal{C}$, $\mu_i$, such that
for all $i,j$ it holds that $\mu_i(\{a_j\}) \in \{1,2,\ldots,M\}$.

We focus on the regime where the number of players is small, so that
$n \gg  M^k$, and in particular $Mk \cdot M^k/n <\epsilon$ for some $\epsilon$. Then it also
holds that $Mk\cdot M^k/\mu_i(\mathcal{C})<\epsilon$.

\subsection{Truthful fair division}
Let $\nu_1,\ldots,\nu_k$ be the set of declared measures.
\begin{definition}
Let $S=(s_1,\ldots,s_k) \in \{1,2,\ldots,M\}^k$
be some vector. Let the bin $B_S \subseteq \mathcal{C}$ be the set of goods
$a$ for which, for each player $i$, it holds that $a \in B_S$ iff
 $\nu_i(a)=s_i$:
\begin{equation}
  \label{eq:bins}
  B_S=\{a \in \mathcal{C} \mbox{ s.t. } \forall i: \nu_i(a)=s_i\}
\end{equation}
Let $\mathcal{B}$ be the set of bins.
\end{definition}

We propose the following mechanism:
\begin{mechanism}
  \label{mech:discrete_k}
 For each bin $B_S$, pick from the
uniform distribution a partition of
it into $k$ parts of equal size $B_S^1,\ldots,B_S^k$, with perhaps some left
over elements which number at most $k-1$.
Let  $C_i'=\bigcup_{B_S \in \mathcal{B}}B_S^i$ and give $C_i'$ to player $i$.
Then, give each leftover good to some player, uniformly at random.
\end{mechanism}
Denote by $C_i$ the set that player $i$ got, i.e. $C_i'$ union any leftovers
given to player $i$. Then it is easy to see that this mechanism is truthful, since
player $i$'s expectation is $\mu_i(\mathcal{C})/k$, independently of her
declared measure $\nu_i$:
\begin{proposition}
$$  \E[\mu_i(C_j)]= \mu_i(\mathcal{C})/k$$
\end{proposition}
\begin{proof}
  This follows from the fact that every $a_l$ ends up in $C_j$ with
  probability $1/k$.
\end{proof}

Truthfulness, however, could have been more simply achieved by, for example,
giving each player the entire set $\mathcal{C}$ with probability $1/k$. This
mechanism's merit
is that it ensures low risk for truthful players:
\begin{theorem}
  When $\nu_i=\mu_i$ then for all $j$ it holds that
  $\mu_i(C_j) \geq (1-\epsilon)\mu_i(\mathcal{C})/k$.
\end{theorem}
when $i=j$ this implies low risk for truthful players.
\begin{proof}
  By definition of the $C_j$'s
  \begin{eqnarray*}
    \mu_i(C_j) \geq \mu_i(C_j')&=& \mu_i\left(\bigcup_{B_S \in \mathcal{B}}B_S^j\right).
  \end{eqnarray*}
  Since the different $B_S^i$'s are disjoint, and by the definition of $B_S$ 
  \begin{eqnarray*}
    \mu_i(C_j) \geq \sum_{B_S \in \mathcal{B}}\mu_i(B_S^j) \geq \sum_{B_S \in \mathcal{B}}s_i|B_S^j|.
  \end{eqnarray*}
  We denote the number of left over
  elements $r_S$, so that $|B_S|=k|B_S^i|+r_S$ for all $i$. Then
  \begin{eqnarray*}
    \mu_i(C_j) &\geq& {\mu_i(\mathcal{C})\over k}-{1\over k}\sum_{B_S \in \mathcal{B}}s_ir_S,
  \end{eqnarray*}
  since $r_S<k$ and $s_i \leq M$, and by the definition of $\epsilon$
  we finally have that
  \begin{equation}
    \label{eq:division_bound}
    \mu_i(C_j) \geq {\mu_i(\mathcal{C}) \over k}-{Mk \cdot M^k\over k} \geq (1-\epsilon)\mu_i(\mathcal{C})/k.
  \end{equation}

\end{proof}

We conclude that assuming players are averse to risk, they may find
actual advantage in playing truthfully, since that will result in a
utility that is with probability one greater than
$(1-\epsilon)\mu_i(\mathcal{C})/k$. Other strategies, on the other
hand, may run the risk of resulting in lower utility.

\subsection{Truthful super-fair division}
We can naturally adapt mechanism~\ref{mech:kepsilon} to the discrete case,
by letting $D$ be the uniform (for example) distribution over the partitions
of  $\mathcal{C}=\{a_1,\ldots,a_n\}$ into $k$ subsets. We then use what is essentially the same
mechanism:
\begin{mechanism}
  \label{mech:discrete_keps}
  Pick a random partition from $D$, keep it iff everyone was allocated strictly
more than $1/k$, and otherwise give everyone $1/k$ using
mechanism~\ref{mech:discrete_k}.
\end{mechanism}

In this discrete case it is easy to see
that if super-fair divisions exist then they are picked with positive
probability. The proofs that this mechanism results in expected utility larger
than $1/k$, and that it is truthful, are identical to the ones for the
continuous case, \ref{prop:k_eps_works} and   \ref{prop:k_eps_truth}.

\subsection{Extending continuous fair division existence results}

Let $\mathcal{M}$ be the space of $k$-by-$n$ matrices $M$ such that
for measures $\mu_1,\ldots,\mu_k$ and some division $C_1,\ldots,C_n$
it holds that $M_{ij}=\mu_i(C_j)$.  Dubins and Spanier
\cite{DubinsSpanier:61}, using a theorem of Lyapunov
\cite{Lyapunov:40}, prove that $\mathcal{M}$ is compact and convex
when the measures $\mu_i$ are non-atomic.  From this follow a plethora
of existence theorems for partitions of different characteristics. For
example, as mentioned above, this fact can be used to show that there
exists a division where each of the $k$ players gets a share worth
$1/k$ by everyone's measure (for probability measures). It can also be
used to show that some players have different measures then a division
exists in which every player gets more than $1/k$, by her own measure.

This result obviously does not apply to the discrete case; the set of
partitions is finite and it is difficult to speak of
convexity. Accordingly, in the general case no fair partition exists,
and a super-fair partition may not exist even when the preferences are
different. A simple example of two goods and three players suffices to
illustrate this point.

One could, however, imagine that all this {\em could} be achieved if
the players were somehow able to share the goods. In fact, if we
allow, for example, that one player has a third of a good and another
two thirds of it (with appropriate utilities), then Dubins and
Spanier's results apply again, and a wealth of partitions with
different qualities is possible again. We refer to this as the
continuous extension of the discrete problem.

However, indivisible goods must by nature remain indivisible. To
overcome this, we propose a randomized partition, similar to the one
used in mechanism~\ref{mech:discrete_k}, that makes possible, {\em in
  expectation}, any partition values possible in the continuous
extension. Moreover, it assures that each player not only receives the
correct utility in expectation, but that in the worst case she will
not receive less than $1-\epsilon$ of what she expects.

We thus consider again a set of indivisible goods
$\mathcal{C}=\{a_1,\ldots,a_l\}$, $k$ players, and their additive
measures $\mu_i$, where $\mu_i(\{a_j\}) \in \{1,2,\ldots,M\}$. We now
imagine that each good can be continuously subdivided, and so define
$a_j^*$ to be copy of $[0,1] \in \R$, and let
$\mathcal{C^*}=\{a_1^*,\ldots,a_l^*\}$. Let $\mathfrak{F}$ be the
standard Borel $\sigma$-algebra on $\mathcal{C}^*$, and let $\nu$ be
the Lebesgue measure on $\mathfrak{F}$.  Define $\mu^*_i$, a measure
on $\mathfrak{F}$, as a continuous extension of $\mu_i$ by
$$\mu_i^*(A)=\sum_j\mu_i(\{a_j\}) \cdot \nu(A\cap a_j^*),$$
for any $A \in \mathfrak{F}$. We refer to $\mathcal{C}^*$ and
$\mu_i^*$ as the {\em continuous extension} of $\mathcal{C}$ and
$\mu_i$.

We are now ready to state the main result of this section:
\begin{theorem}
  Consider the problem of partitioning indivisible goods as defined
  above, and its continuous extension.  Let $\mathcal{M}$ be the space
  of $k$-by-$n$ matrices $M$ such that for some $C_1^*,\ldots,C_n^*$
  it holds that $M_{ij}=\mu_i^*(C_j^*)$.

  Then for every element $M \in \mathcal{M}$ there exists a randomized
  partition $C_1,\ldots,C_n$ satisfying $\E[\mu_i(C_j)] = M_{ij}$,
  and moreover
  \begin{equation*}
    {\mu_i^*(C_j^*)\over \mu_i^*(\mathcal{C}^*)}+k\epsilon
    \geq {\mu_i(C_j) \over \mu_i^*(\mathcal{C}^*)}
    \geq {\mu_i^*(C_j^*)\over \mu_i^*(\mathcal{C}^*)}-\epsilon,
  \end{equation*}
  where $\epsilon$, as before, is $O(M^k/n)$.
\end{theorem}
\begin{proof}
  Given a division $C_1^*,\ldots,C_n^*$ of the divisible $C^*$, we
  would like to divide the discrete $\mathcal{C}$ in a way that
  approximates this division as closely as possible. That is, we would
  like to find a division $C_1,\ldots,C_n$ such that
  $\mu_i(C_j)\approx \mu_i^*(C_j^*)$. We propose two schemes to do
  this: the random scheme and the binned scheme.  For both of them, we
  define an $n$-by-$l$ matrix $D$ ($l$ being the number of indivisible
  goods), where $D_{ij}$ is the fraction of good $a_j^*$ that belongs
  to $C_i^*$:  $D_{ij}=\nu(C_i^*\cap a_j^*)$.

  \paragraph{The random scheme.}

    In the random scheme, we simply give player $i$ good $a_j$ with
    probability $D_{ij}$ (note that by the definition of $D$,
    $D_{i\cdot}$ is a distribution). Then
    \begin{eqnarray*}
      \E[\mu_i(C_j)] &=& \sum_m\mu_i(\{a_m\}) \cdot \P[a_m \in C_j]
      = \sum_m\mu_i(\{a_m\}) \cdot D_{jm}
      = \mu_i^*(C_j^*)
    \end{eqnarray*}
    and its standard deviation is
    $O\left(\sqrt{\mu_i^*(C_j^*)}\right)$.

  \paragraph{The binned scheme.}
    In the binned scheme, we bin the elements of $\mathcal{C}$ into
    bins $\{B_S\}$ as above. Without loss of generality, let there be,
    for each player $i$ and bin $B_S$, a single value $D_{iS}$ such
    that $D_{im}=D_{iS}$ for all $m$. No generality is indeed lost:
    because all elements of a bin are equivalent to all the players,
    then for any partition $C_j^*$ there exists an equivalent
    partition, in the sense of the utilities of the players, for which
    such $D_{iS}$'s exist.

    Let $n_S$ be the number of elements in bin $B_S$. From each bin
    $B_S$, we give player $i$ a number of goods equal to $\lfloor
    n_SD_{iS}\rfloor$, picked from the uniform distribution over such
    partitions. Any leftover $a_m$ we give according to the random
    scheme, i.e. to player $i$ with probability $D_{iS}$.

    The expectation for $\mu_i(C_j)$ clearly remains
    $\mu_i^*(C_j^*)$. However, here we can bound it from below:
    \begin{equation*}
      {\mu_i(C_j) \over \mu_i^*(\mathcal{C}^*)}
      \geq {1\over \mu_i^*(\mathcal{C}^*)}\sum_{B_S\in \mathcal{B}}(n_SD_{iS}-1)s_i
      \geq {\mu_i^*(C_j^*)\over \mu_i^*(\mathcal{C}^*)}-\epsilon.
    \end{equation*}
    We can also bound it from above, by noting that the most a player
    can get beyond $\mu_i(C_j)$ is what's lost by the rest of the
    players:
    \begin{equation*}
      {\mu_i^*(C_j^*)\over
        \mu_i^*(\mathcal{C}^*)}+k\epsilon \geq {\mu_i(C_j) \over
        \mu_i^*(\mathcal{C}^*)}
    \end{equation*}
\end{proof}

\appendix
\section{Existence of a distribution on partitions with positive probability
for super-fair division}
Let $\mathcal{R}$ be the set of finite unions of half-open intervals of $[0,1)\in \R$
with rational endpoints. Let $\mathcal{Q}$ be defined to be the
class of partitions of $[0,1)\in \R$ such that each part of the partition
is in $\mathcal{R}$.

The main lemma we want to prove is the following:
\begin{lemma}
  Let $\mu_1,\ldots,\mu_k$ be non-atomic Borel measures on $[0,1)\in \R$ and the Borel $\sigma$-algebra
$\mathcal{B}$. Let $B_1,\ldots,B_k$ be a partition of the interval, where $B_i \in \mathcal{B}$ and $\delta > 0$.
Then there exists a partition $Q_1,\ldots,Q_k$ in $\mathcal{Q}$
such that $\mu_i(B_i \triangle Q_i) < \delta$ for all $i$.
\end{lemma}

\begin{proof}
The proof uses the fact that $\mathcal{Q}$ is an algebra, i.e., it is closed under finite unions, intersections and taking of complements.
Let $\delta'$ be chosen later. Since all of the measures are Borel we can find open
sets $O_{i,j}$ so that $B_j \subset O_{i,j}$ and
$\mu_i(O_{i,j} \triangle B_j) < \delta'$ (see, e.g., theorem 2.17 in \cite{Rudin:87}).
Taking $O_j = \cap_i O_{i,j}$, we get open sets such that
$B_j \subset O_j$ and $\mu_i(B_j \triangle O_j) < \delta'$ for all $i$ and $j$.

Fix $j$ and note that $O_j = \cup_{n=1}^{\infty} I_{j,n}$, where $I_{j,n}$ are open intervals with rational
end-points.
Take $m$ sufficiently large so that $\mu_i(O_j \setminus \cup_{n=1}^m I_{j,n}) < \delta'$ for all $i$.
Let $J_{j,n}$ be the same as $I_{j,n}$ except that the left-end point of the interval is
added and let
$\tilde{P}_j = \cup_{n=1}^m J_{j,n}$. Since the measures are non-atomic we have $\mu_i(\cup_{n=1}^m J_{j,n}) > \mu_i(O_j) - \delta'$.
Note that the $\tilde{P}_j$'s are all unions of half-open intervals with rational end-points.
Moreover, for all $i$ and $j$,
\[
\mu_i(\tilde{P}_j \triangle B_j) \leq \mu_i(\tilde{P}_j \triangle O_j) + \mu_i(O_j \triangle B_j) \leq 2 \delta'.
\]
Note further that for all $i$ it holds that:
\[
\mu_i[0,1) \geq \mu_i(\cup_j \tilde{P}_j) \geq \mu_i(\cup_j O_j) - \sum_j \mu_i(O_j \setminus \cup_{n=1}^m I_{j,n}) \geq \mu_i[0,1) - k \delta',
\]
so
\[
\mu_i([0,1) \setminus \cup_j \tilde{P}_j) \leq k \delta'.
\]
Now take $P_i = \tilde{P_i}$ for $i > 1$ and $P_1 = \tilde{P}_1 \cup ([0,1) \setminus \cup_j \tilde{P}_j)$.
Then $\cup P_i = [0,1)$ and 
\[
\mu_i(P_j \triangle B_j) \leq (k+2) \delta'
\]
for all $i$ and $j$. 
The $P_i$ are almost the desired partition. They satisfy all the needed
properties except that they are not a partition. We now take $Q_j = P_j \setminus \cup_{j' < j} P_{j'}$.
$Q_j$ is obviously a partition. Moreover:
\begin{eqnarray*}
\mu_i(Q_j \triangle B_j) &\leq& \mu_i(P_j \triangle B_j) + \sum_{j' \neq j} \mu_i(P_{j'} \triangle B_j)
\\ &\leq& \mu_i(P_j \triangle B_j) + \sum_{j' \neq j} \mu_i(B_{j'} \triangle B_j) + \sum_{j' \neq j} \mu_i(P_{j'} \triangle B_{j'})
\\ &\leq& 2 k (k+2) \delta'.
\end{eqnarray*}
Taking $\delta' = \delta/(2 k (k+2))$ concludes the proof. 
\end{proof}

\begin{theorem}
  Let $\mu_1,\ldots,\mu_k$ be non-atomic probability measures on
  $[0,1) \in \R$ and the Borel $\sigma$-algebra,
  such that there exist $i,j$ for which $\mu_i \neq \mu_j$. Let $D$
  be a distribution over the partitions $C_1,\ldots,C_k$ of $[0,1)$ into $k$
  sets, such that the support of $D$ is $\mathcal{Q}$. Then
  \begin{equation*}
    \P_D[\forall i:\:\mu_i(C_i)>1/k]>0.
  \end{equation*}
\end{theorem}

\begin{proof}
  \label{app:kepsilon_proof}
  By Dubins and Spanier's theorem, there exists a partition
  $B_1,\ldots,B_k$ of measurable sets such that for all $i$ it holds that
  $\mu_i(B_i)>1/k$. Let $\epsilon>0$ be such that $\mu_i(B_i)>1/k+\epsilon$.

  By the lemma above there exists a partition
  $Q_1,\ldots,Q_k$ in $\mathcal{Q}$
  such that
  $\forall i,j:\:\mu_i(Q_j \triangle B_j)<\half\epsilon$. This, in particular,
  implies for all $i$ that $\mu_i(Q_i)>1/k+\half\epsilon$.
\end{proof}

\bibliographystyle{abbrv} \bibliography{cake}
\end{document}